\documentclass[a4paper,12pt]{article}
\usepackage{amsmath,amssymb}
\usepackage{graphicx}
\usepackage{mathrsfs}
\usepackage{float}
\usepackage{ulem}

\newcommand{\be}{\begin{equation}}
\newcommand{\ee}{\end{equation}}
\newcommand{\f}{\frac}

\newcommand{\p}{\partial}

\newcommand{\bea}{\begin{eqnarray}}
\newcommand{\eea}{\end{eqnarray}}
\newcommand{\ba}{\begin{align}}
\newcommand{\ea}{\end{align}}

\newcommand{\la}{\langle}
\newcommand{\ra}{\rangle}

\begin{document}
\begin{titlepage}

\vspace{.4cm}
\begin{center}
\noindent{\Large \textbf{Perturbative expansions of  R\'enyi relative divergences and holography}}\\
\vspace{1cm}

 Tomonori Ugajin

\vspace{.5cm}
  {\it
 Okinawa Institute of Science and Technology, \\
Tancha, Kunigami gun,  Onna son, Okinawa 1919-1 \\
\vspace{0.2cm}
 }
\end{center}


\begin{abstract}
In this paper, we develop a novel way to perturbatively calculate R\'enyi relative divergences  $D_{\gamma}(\rho|| \sigma) ={\rm tr} \rho^{\gamma} \sigma^{1-\gamma}$ and related quantities without using replica trick as well as analytic continuation. We explicitly determine  the form of the perturbative term at any order by an integral along the modular flow of the  unperturbed state. By applying the prescription to a class of  reduced density matrices in conformal field theory, we  find that the second order term of certain linear combination of the divergences  has a holographic expression in terms of 
bulk symplectic form, which is a one parameter generalization of  the statement "Fisher information = Bulk canonical energy".  
\end{abstract}

\end{titlepage}

\tableofcontents

\section{Introduction}

The concept of entanglement is one of the keys to understand how holography works. This idea is supported by the Ryu Takayanagi formula \cite{Ryu:2006bv,Ryu:2006ef} and its covariant generalization \cite{Hubeny:2007xt}, which relate the area of particular  extremal surfaces  in the bulk,  to  the entanglement entropies in the dual  conformal field theory (CFT).  As a concrete and quantitative application of this entanglement vs gravity program,  recently it has been shown that  bulk gravitational dynamics can be read off  from the entanglement structure of states in the dual CFT.   

\vspace{0.2cm}

In this line of developments, it was initially observed that so called first law of entanglement \cite{Bhattacharya:2012mi} is related to  the linearized Einstein equations in the bulk \cite{Lashkari:2013koa,Faulkner:2013ica}. Consider starting from the vacuum reduced density matrix $\rho_{0}$ and making it excited slightly  $\rho_{0} \rightarrow \rho_{0} +\delta \rho$ in a CFT. The  change of the entanglement entropy $\delta S$ obeys  first law of entanglement,  $\delta S ={\rm tr} \left[ K \delta_{0} \right]$, where $K =-\log \rho_{0}$ is called modular Hamiltonian of $\rho_{0}$.  For the subsystems  of special type, the vacuum modular Hamiltonian has a local expression given by an integral of energy density over the subsystem.  There is a natural bulk counterpart of the vacuum modular Hamiltonian, namely, the generator of  time translation  of a topological black hole with hyperbolic horizon, whose Bekenstein Hawking entropy gives the CFT vacuum entanglement entropy \cite{Casini:2011kv}. The first law of entanglement is related to the first law of thermodynamics applied to the topological black hole, and this enabled us to read off the linearized equations of motion.

\vspace{0.2cm}

Recently this nice story at the  first order in the perturbation  $\delta \rho$ has been generalized to the  quadratic order .  It was noticed that, in CFT the  second order change of the entanglement entropy  can be concisely summarized  as an integral of correlation functions along the flow generated by the vacuum modular Hamiltonian $K =-\log \rho_{0}$ on the subsystem \cite{Faulkner:2014jva,Faulkner:2015csl,Faulkner:2017tkh}. This was further extended to arbitrarily order in  $\delta \rho$ and some technical issue was pointed out \cite{Sarosi:2017rsq}. It was also recognized that  by rewriting the CFT answer in terms of bulk variables, we naturally identify it with bulk canonical energy\cite {Hollands:2012sf}, which was first found holographically in \cite{Lashkari:2015hha}.  This makes it possible 
to read off the bulk equations of motion beyond the linearized level.

\vspace{0.2cm}

Given these developments, it is now natural to generalize this story to  other quantum information theoretic quantities.  In particular we would like to find  such a quantity which admits a nice perturbative expansion in  CFT   and has a  dual holographic expression.  Natural candidates having these properties  are those involving  powers of reduced density matrices, for example ${\rm tr}\;  \rho^{\gamma}$ which is related to  R\'enyi entropy.

\vspace{0.2cm}

Conventionally, a  R\'enyi type quantity, like  ${\rm tr}\;\rho^{\gamma}$ has been computed by 
replica trick. In this trick,  we first regard  the   R\'enyi  index to be a positive integer $\gamma =n$, and represent the quantity as a path integral on a branched space $\Sigma_{n}$ which is prepared by gluing $n$ copies of the original space with cuts along the subsystems. After the computation of the path integral, 
we then  analytically continue the integer $n$ to arbitrarily number $\gamma$. However,  this trick has several 
disadvantages, even when we compute the quantity perturbatively.   First of all, the analytic continuation is usually difficult to perform.  For example,  when we  perturbatively  expand   ${\rm tr}\;  \rho^{n}$ for $\rho=\rho_{0} +\delta \rho$,  at quadratic order we encounter  following sum

\be  
\sum_{k,m} {\rm tr} \left[\rho_{0}^{k-1} \delta \rho \rho_{0}^{m-k-1}  \delta \rho \rho_{0}^{n-m} \right]. \label{eq:brute}
\ee 
 
In order to analytically continue it in $n$ we first need to perform  this sum to get  a closed expression.  Although for special cases we can do this, in general it is difficult. In addition to this, we do not know how to do analogous  sums for  the cubic term and higher.   Second, there are ambiguities in the analytic continuations. According to the Carson's theorem, we need to correctly specify the behavior of ${\rm tr} \; \rho^{n}$ on certain region of the complex $n$ plane, in order to fix the ambiguities.

\vspace{0.2cm}
%
%

 In order to overcome these difficulties, in this paper we would like to  develop a new way to perturbatively calculate R\'enyi  type quantities without using replica trick,  and analytic continuation. The idea we employ is simple, 
namely writing  ${\rm tr}\;\rho^{\gamma}$ by a contour integral, 
\be 
{\rm tr}\;\rho^{\gamma} =\int_{C}\f{dz}{2\pi i} \;  z^{\gamma}\;  {\rm tr} \f{1}{z-\rho }, \label{eq;resolvent}
\ee
where the contour $C$ is chosen so that it includes all the poles of the integrand, but avoid the contribution of the branch cut coming from $z^{\gamma}$. 
 We refer to \cite{Casini:2009sr, Witten:2018lha} for discussions on the representation. By expanding the denominator of the integrand for perturbative states  $\rho=\rho_{0} +\delta \rho$, we can systematically write each  term of the perturbative expansion by an integral along the modular flow of the reference state $\rho_{0}$. If we apply this expansion for a class of perturbative excited states from  vacuum  in a $d$ dimensional CFT, we can write each term as an integral of a correlation function $ \la \cdots \ra_{\Sigma_{\gamma}} $on the branched space $\Sigma_{\gamma} = S^{1}_{\gamma} \times H^{d-1}$ along the modular flow generated by $\rho_{0}$.  
Here,  $S^{1}_{\gamma} $ denotes the Euclidean time circle with $ 2\pi \gamma$ periodicity, and $H^{d-1}$
is $d-1$ dimensional hyperbolic space.  Of course, the CFT correlation functions on $\Sigma_{\gamma}$ are difficult to calculate when $d>2$, 
 as the branched space  $\Sigma_{\gamma}$ is not conformally related to d dimensional flat space, and  even two point functions  are highly theory dependent ones.

\vspace{0.2cm}

However, by the same trick, we can similarly  expand the Petz's quasi entropy \cite{1985787} defined by,
\be 
D_{\gamma} (\rho||  \sigma) = {\rm tr} \; \ \rho^{\gamma}\sigma^{1-\gamma} .
\ee

This quantity can be regarded as a one parameter generalization of relative entropy, 
\be 
\f{d}{d \gamma}D_{\gamma} (\rho||  \sigma) \big|_{\gamma=1} =S(\rho||\sigma) =  {\rm tr}  \rho \log \rho -{\rm tr} \rho \log \sigma.
\ee

We also refer to recent studies on R\'enyi generalizations of relative entropy \cite{Lashkari:2014yva,Seshadreesan:2017akv,Casini:2018cxg,Lashkari:2018nsl,Bernamonti:2018vmw,May:2018tir} as well as perturbative calculations of relative entropy \cite{Blanco:2013joa,Sarosi:2017rsq,Sarosi:2016oks,Sarosi:2016atx,Ugajin:2016opf,Nakagawa:2017fzo, Takayanagi:2018zqx,Lashkari:2018oke}.
One notable feature of  this R\'enyi relative divergence is that, each term of  its  perturbative expansion  involves a correlator on the regular space $\Sigma_{1}$ which is conformally related to flat space.  This implies that the first few terms of  the expansion are almost fixed  by conformal symmetry, and  independent of the CFT we consider.  In particular,  this property enables us to holographically write the 
quadratic terms of certain linear combinations of $D_{\gamma} (\rho||  \sigma)$ which we will denote by $X_{\gamma}(\delta \rho),  Y_{\gamma}(\delta \rho)$, in terms of bulk symplectic form, without the details of the bulk to boundary dictionary.  This generalizes the statement " quantum fisher information = bulk canonical energy".  See also \cite{Belin:2018fxe, Belin:2018bpg} for recent discussions on bulk symplectic form.

This paper is organized as follows. In section \ref{sec;new}, we explain how to expand $T_{\gamma} (\rho)={\rm tr} \rho^{\gamma}$ using the formula (\ref{eq;resolvent}). We first derive expressions of the perturbative terms as integrals with respect to the entanglement spectrum of the unperturbed state. In section \ref{sec;checks}, we check these expressions against known results. In section \ref{sec;modflow} we express each term of the perturbative expansion as an integral along the modular flow of the unperturbed state by Fourier transforming the  spectral representation of the kernel derived in   section \ref{sec;new}.  In section \ref{sec;cft} we apply the formalism to  reduced density matrices in conformal field theory, and write these perturbative terms in terms of correlation functions in CFT.  In section \ref{sec;holographic}, we discuss a similar expansion of Petz's quasi entropy and  derive a holographic expression of the second order term.

\section{New expansion formula using the resolvent trick}  
\label{sec;new}

In the first few sections we focus on the R\'enyi type quantity
\be
T_{\gamma}(\rho) = {\rm tr} \; \rho^{\gamma} .
\ee

In the discussions we do {\it not} assume the index $\gamma$ to be an positive integer $\gamma \in \mathbb{Z}_{+}$, where one can use the replica trick. Although we will apply the prescription developing here to conformal field theory, the discussions in this  section  and the next few  ones are applicable for any density matrix of any theory.  

When the density matrix $\rho$ is sufficiently close to the reference state $\rho_{0}$, ie $\rho=\rho_{0}+\delta \rho$,  we can expand $T_{\gamma}(\rho)$ by a power series of  $\delta \rho$,

\be 
T_{\gamma} (\rho) =T_{\gamma} (\rho_{0}) + \sum^{\infty}_{n=0}T^{(n)}_{\gamma}(\delta \rho), \label{eq;expansion}
\ee

and  decompose  each term in the perturbative expansion by the spectra of the reference state $\rho_{0}$. 
Let us first do this.

\begin{figure}
\begin{center}
\includegraphics[width=8cm]{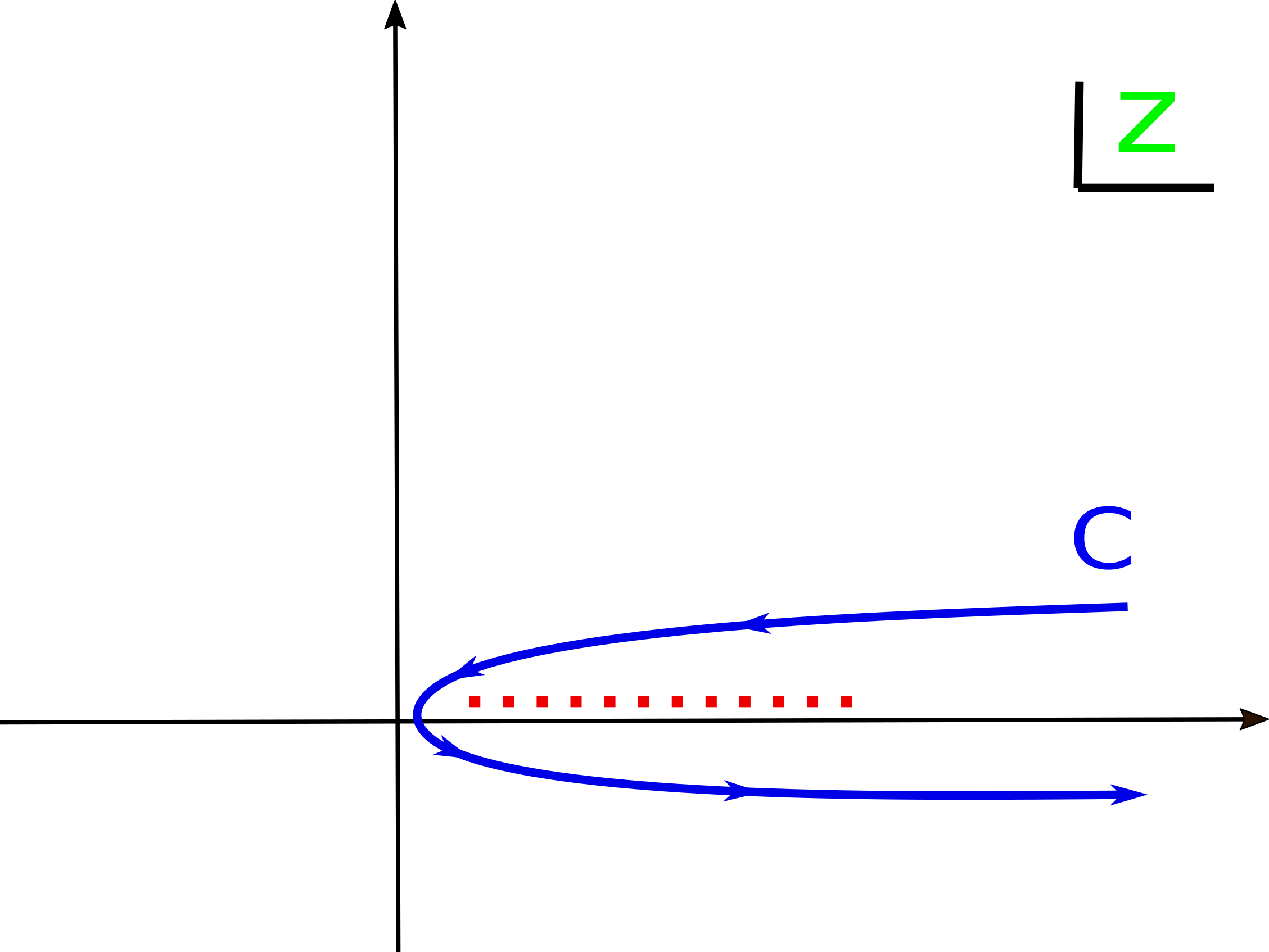}
\end{center}
\caption{The contour  $C$ of the integral (\ref{eq:resol})(blue line). Red dots are the poles of the meromorphic function $f(z)={\rm tr}\; \f{ 1}{z-\rho }$. These poles are in the segment $0<{\rm Re} z<1$. }
\label{fig;zcontour}
\end{figure}

We begin the discussion by first writing  $T_{\gamma}(\rho)$ using the  resolvent of $ \rho$, 
\be 
{\rm tr} \; \rho^{\gamma} = \int_{C}\f{dz}{2\pi i} \;  z^{\gamma}\;  {\rm tr} \f{1}{z-\rho }, \label{eq:resol}
\ee
 
where the contour $C$ is encircling the interval $[\rho_{\min},1]$ in the $z$ plane, but not $z=0$, so that it picks up all contributions of eigenvalues of $\rho$.  $\rho_{\min} $ is the smallest eigenvalue of 
the density matrix $\rho$.(See figure \ref{fig;zcontour}.)
 When $\rho$ is a reduced density matrix of  a quantum field theory, we need to put a UV cut off  $\varepsilon$ so that the density  matrix $\rho$ has a minimum eigenvalue, then after the calculation we send  $\varepsilon \rightarrow 0$. We will explicitly see that only  the unperturbed  term $T_{\gamma} (\rho_{0})$ depends on the UV cutoff and rests do not. Therefore we can uniquely fix the the form of $T^{(n)}_{\gamma}(\delta \rho),\; n\geq 1$.

When $\rho= \rho_{0} +\delta \rho$ the resolvent can be easily expanded,

\be
 \f{1}{z-\rho} = \sum^{\infty}_{n=0} R_{n}(\delta \rho) \quad R_{n} (\delta \rho) =\left(\f{1}{(z-\rho_{0})} \delta \rho  \right)^{n} \f{1}{(z-\rho_{0})}.
\ee

%

By inserting the complete set of  eigenstates $|\omega_{i} \ra$ of the reference state $\rho_{0}$ ,
\be 
\int d \omega_{i} | \omega_{i} \rangle \langle \omega_{i}  |  =1, \quad \rho_{0} | \omega_{i} \rangle=e^{-2\pi \omega_{i}} |\omega_{i} \rangle, \label{eq;specdec}
\ee

to the left of $i$-th term,   taking trace,  and  evaluating $1/(z-\rho_{0})$ from the left, we have, 



\begin{align} 
{\rm tr} \left[R_{n} (\delta \rho)\right] &=\int \prod^{n}_{i=1}d \omega_{i} \prod^{n-1}_{i=1} \f{1}{z-e^{-2\pi \omega_{i}}} \prod^{n-1}_{k=1}  \langle \omega_{k}|\delta \rho | \omega_{k+1} \rangle \langle  \omega_{n}| \f{1}{(z-\rho_{0})}  \delta \rho \f{1}{(z-\rho_{0})}  |\omega_{1} \rangle \nonumber \\
&=\int \prod^{n}_{i=1}d \omega_{i}\f{1}{(z-e^{-2\pi \omega_{1}})^{2}}\prod^{n}_{i=2} \f{1}{z-e^{-2\pi \omega_{i}}} \prod^{n}_{k=1}  \langle \omega_{k}|\delta \rho | \omega_{k+1} \rangle ,
\end{align}  

in the last term,  $\omega_{n+1}\equiv \omega_{1}$ is understood.

In summary,  here we expanded $T_{\gamma} (\rho)$ with respect to $\delta \rho$, as in  (\ref{eq;expansion}), and saw that the $n$ th order term of the expansion $T^{(n)}_{\gamma}(\delta \rho)$ is given by

\be 
T^{(n)}_{\gamma}(\delta \rho) = \int  \prod^{n}_{i=1} d\omega_{i} \left[\int_{C} \f{dz}{2\pi i} z^{\gamma} \f{1}{(z-e^{-2\pi \omega_{1}})^{2}}\prod^{n}_{i=2} \f{1}{z-e^{-2\pi \omega_{i}}}\right] \prod^{n}_{k=1}  \langle \omega_{k}|\delta \rho | \omega_{k+1} \rangle. 
\ee

By defining the kernel function,
\be
K^{(n)}(\omega_{1}, \cdots \omega_{n})  \equiv \int_{C} \f{dz}{2\pi i}  \f{z^{\gamma}}{(z-e^{-2\pi \omega_{1}})^{2}}\prod^{n}_{i=2} \f{1}{z-e^{-2\pi \omega_{i}}},
\ee

we write, 
\be
T^{(n)}_{\gamma}(\delta \rho) = \int  \prod^{n}_{i=1} d\omega_{i}  K^{(n)}(\omega_{1}, \cdots \omega_{n})\prod^{n}_{k=1}  \langle \omega_{k}|\delta \rho | \omega_{k+1} \rangle. \label{eq;tnes}
\ee


\section{ Some explicit checks} 
\label{sec;checks}

We have obtained the perturbative expansion using the spectrum  of the reference state $\rho_{0}$. To get some insights, in this section  we explicitly write down first few terms of the expansion and check them against known results.

\subsection{ First order term $T^{(1)}_{\gamma}(\delta \rho)$} 

The first order term of the series  is given by 
\begin{align} 
T^{(1)}_{\gamma}(\delta \rho) &=\int d\omega \;  \la \omega |\delta \rho | \omega \ra \int_{C} \f{dz}{2\pi i} \f{z^{\gamma}}{(z-e^{-2\pi \omega})^2} \nonumber \\[+10pt]
&= \gamma {\rm tr}  \left[\rho_{0}^{\gamma-1} \delta \rho \right].
\end{align}
as it should be. 

\subsection{Second order term   $T^{(2)}_{\gamma}(\delta \rho)$}

Let us move on to the second order term $T^{(2)}_{\gamma}(\delta \rho)$. It is given by 

\begin{align} 
 T^{(2)}_{\gamma}(\delta \rho) =  
\int d\omega d\omega' \la \omega |\delta \rho | \omega' \ra  \la \omega' |\delta \rho | \omega \ra \;  K (\omega, \omega'). \label{eq:renyi2}
\end{align}


Precise form of $K (\omega, \omega')$ can be derived by the contour integral, 

\begin{align} 
K (\omega, \omega') &= \int_{C} \f{dz}{2\pi i} \; \f{z^{\gamma} }{(z-e^{-2\pi \omega})^2 (z-e^{-2\pi \omega'})} \nonumber \\[+10 pt]
&= \f{1}{(e^{-2\pi\omega'}-e^{-2\pi\omega})^2}\left[ (\gamma-1) e^{-2\pi \gamma\omega} +e^{-2\pi \gamma\omega'} - \gamma e^{-2\pi (\gamma-1) \omega}e^{-2\pi \omega'}  \right]. \label{eq:kfreaq}
\end{align}


\subsubsection{Checks} 


\underline{$\gamma=n \in \mathbb{Z}_{+}$} 


When the index $\gamma$ is a positive integer,  the kernel $K_{\gamma} (\omega, \omega') $ is  decomposed into the sum, 
\begin{align} 
K (\omega, \omega') = \left[ \sum^{\gamma-2}_{l=0} \left( (\gamma-1) -l\right) \left(e^{-2\pi \omega} \right)^{\gamma-l}  \left(e^{-2\pi\omega'} \right)^{l} \right] .
\end{align}

Plugging this into (\ref{eq:renyi2}) and undoing the spectral decomposition, we recover the obvious expansion (\ref{eq:brute}) which we frequently encounter in replica calculations. The kernel avoids the difficulties of  replica trick, by automatically doing the summation as well as analytic continuation in $n$. 

\hspace{0.1cm}

\underline{The von Neumann entropy limit}  


$T_{\gamma} (\rho)$ is related to the von Neumann entropy $S(\rho)$ by 
\be 
S(\rho) = -{\rm tr} \rho \log \rho = \f{\p}{\p \gamma} T_{\gamma} (\rho) \big|_{\gamma=1} .
\ee

From (\ref{eq:kfreaq}) we derive the kernel for the quadratic part of the von Neumann entropy,
\be 
\f{\p K_{\gamma}}{\p \gamma} \big|_{\gamma=1} = \f{e^{2\pi\omega}}{(1-e^{2\pi(\omega-\omega')} )} \left[(e^{-2\pi\omega} -e^{-2\pi\omega'})  +2\pi(\omega- \omega')e^{-2\pi \omega'} \right] . \label{eq;entk}
\ee

In \cite{Faulkner:2014jva},  a  perturbative expansion of  the von Neumann entropy $S(\rho_{0} +\delta \rho)$ was discussed, by expanding the modular Hamiltonian $K_{\rho} =- \rho_{0} +\delta \rho$ using the   identity, 
\be 
\log \rho = \int^{\infty}_{0} d \beta \left( \f{1}{\rho+ \beta} -\f{1}{\beta+1} \right),
\ee

the result  of the quadratic order kernel in \cite{Faulkner:2014jva} agrees with (\ref{eq;entk}).

\section{ Expressions of  perturbative terms  in terms of the vacuum  modular flow}
\label{sec;modflow}

The $\omega$ integrals in the right hand side of (\ref{eq;tnes}) are of course hard to perform, as we do not know precise form of the  eigenvalue distribution of $\rho_{0}$. To proceed, we now express each term of the 
perturbative series  $T^{(n)}_{\gamma}(\delta \rho)$ as an integral along   the  modular flow of $\rho_{0}$, by Fourier transforming the kernel $\mathcal{K}^{(n)}_{\gamma}(\omega_{1}, \cdots \omega_{n}) $.

This process is very analogous to the case of the von Neumann entropy perturbation done in \cite{Faulkner:2014jva} for quadratic order term  and generalized to higher order terms in \cite{Sarosi:2017rsq}.  It is convenient to introduce the rescaled  kernel,  defined by

\begin{align} 
\mathcal{K}^{(n)}_{\gamma}(\omega_{1}, \cdots \omega_{n}) &\equiv e^{2\pi \gamma \omega_{1} -2\pi \sum_{k=1}^{n} \omega_{k}}K^{(n)}(\omega_{1}, \cdots \omega_{n}), \\
&=\int_{C} \f{dz}{2\pi i} z^{\gamma}  \f{e^{2\pi (\gamma-1)\omega_{1}}}{(z-e^{-2\pi \omega_{1}})^{2}}\prod^{n}_{i=2} \f{e^{-2\pi \omega_{i}}}{z-e^{-2\pi \omega_{i}}}.
\end{align}

Using this function, we  get
\begin{align}
T^{(n)}_{\gamma}(\delta \rho)&= \int  \prod^{n}_{i=1} d\omega_{i}  K^{(n)}(\omega_{1}, \cdots \omega_{n})\prod^{n}_{k=1}  \langle \omega_{k}|\delta \rho | \omega_{k+1} \rangle \nonumber \\
&= \int  \prod^{n}_{i=1} d\omega_{i} \;  e^{-2\pi \gamma \omega_{1} +2\pi \sum_{k=1}^{n} \omega_{k} }\; \mathcal{K}^{(n)}_{\gamma} (\omega_{1}, \cdots \omega_{n}) \prod^{n}_{k=1}  \langle \omega_{k}|\delta \rho | \omega_{k+1} \rangle \nonumber \\
&=\int  \prod^{n}_{i=1} d\omega_{i} \; \mathcal{K}_{n}^{\gamma}(\omega_{1}, \cdots \omega_{n}) \; \langle \omega_{1} | e^{-2\pi \gamma K}\delta \tilde{\rho} | \omega_{2} \rangle\prod^{n}_{k=1}  \langle \omega_{k}|\tilde{\delta} \rho | \omega_{k+1} \rangle, \nonumber \\
\end{align}

where $2\pi K =-\log \rho_{0}$ is the modular Hamiltonian of $\rho_{0}$, and $\tilde{\delta} \rho= e^{\pi K}\delta \rho \; e^{\pi K}$.
It can be easily shown that the new kernel $\mathcal{K}^{(n)}_{\gamma}(\omega_{1}, \cdots \omega_{n})$  is invariant under the shifts $\omega_{i} \rightarrow \omega_{i} +\alpha$,
\be 
\mathcal{K}^{(n)}_{\gamma}(\omega_{1} +\alpha, \cdots \omega_{n} +\alpha) =\mathcal{K}^{(n)}_{\gamma}(\omega_{1}, \cdots \omega_{n}) .
\ee

So if we change the variables to  $\{ a_{i}, b\}$, 
\be 
a_{i} = \omega_{i} -\omega_{i+1}, i=1 \cdots n-1 \quad b= \sum_{i=1}^{n} \omega_{i}, \label{eq; relation}
\ee
$\mathcal{K}^{(n)}_{\gamma}(\omega_{1}, \cdots \omega_{n})$  only depends on $n-1$ variables  $\{ a_{i} \}_{ i=1 \cdots n-1}$,
\be
\mathcal{K}^{(n)}_{\gamma}(\omega_{1}, \cdots \omega_{n}) =\mathcal{K}^{(n)}_{\gamma}(a_{1}, a_{2}, \cdots a_{n-1}).
\ee

Thanks to this property  $\mathcal{K}^{(n)}_{\gamma}(\omega_{1}, \cdots \omega_{n}) $ has a nice Fourier transformation, 
\be 
\mathcal{K}^{(n)}_{\gamma}(\omega_{1}, \cdots \omega_{n})= \int_{C} ds_{1} \cdots ds_{n-1} e^{i \sum_{k=1}^{n-1} s_{k}a_{k}} \; \mathcal{K}^{(n)}_{\gamma} (s_{1}, \cdots s_{n-1}), \label{eq;fouriesr}
\ee

$\{s_{i} \}_{ i=1 \cdots n-1}$   are variables dual to  the spectrum of $\rho_{0}$, therefore they have a geometric interpretation, ie,  they are parameterizing the modular flow of  $\rho_{0}$.  Also, as we will see later, we  need to properly choose the integration contours $C$ in order for the Fourier transformation  (\ref{eq;fouriesr}) to correctly reproduce the kernel $\mathcal{K}^{(n)}_{\gamma}(\omega_{1}, \cdots \omega_{n})$. 

Using this and undoing the spectral decompositions (\ref{eq;specdec}), we can write $T^{(n)}_{\gamma}(\delta \rho)$ as an integral  of   real time $\{s_{i} \}$ variables,
\begin{align}
T^{(n)}_{\gamma}(\delta \rho)
&=\int  \prod^{n}_{i=1} d\omega_{i}\;  \mathcal{K}^{(n)}_{\gamma} (\omega_{1}, \cdots \omega_{n})  \; \langle \omega_{1} | e^{-2\pi \gamma K}\delta \tilde{\rho} | \omega_{2} \rangle\prod^{n}_{k=1}  \langle \omega_{k}|\tilde{\delta} \rho | \omega_{k+1} \rangle  \nonumber \\
&=\int_{C} ds_{1} \cdots ds_{n-1} \;  \mathcal{K}^{(n)}_{\gamma} (s_{1}, \cdots s_{n-1}) \;  {\rm tr} \left[ e^{-2\pi \gamma K}\prod^{n-1}_{k=1} e^{i Ks_{k}} \tilde{\delta} \rho \;  e^{-i Ks_{k}}  \; \tilde{\delta} \rho \right] . \label{eq;pertfinal}
\end{align}

In the actual CFT computations, this  undoing is a bit tricky, and needed  special cares. We will discuss on this in the latter sections.

\subsection{ Doing the Fourier transformation}

Let us first specify the form of the real time kernel $\mathcal{K}^{(n)}_{\gamma} (s_{1}, \cdots s_{n-1})$.The task is doing the inverse Fourier transformation, 

\be 
\mathcal{K}^{(n)}_{\gamma} (s_{1}, \cdots s_{n-1}) =\int \f{da_{1} \cdots da_{n-1}}{(2\pi)^{n-1}} e^{-i \sum_{k=1}^{n-1} s_{k}a_{k}}\mathcal{K}^{(n)}_{\gamma}(a_{1}, \cdots a_{n-1}).
\ee

The trick we use is very similar to the one developed  in our  previous paper \cite{Sarosi:2017rsq}.  By inserting a delta  function, 
\be 
\delta (q) = \f{1}{2\pi} \int db e^{-iqb},
\ee
we can disentangle the multiple  integral to a product of integrals of single variables $\{ \omega_{i} \}$, 
\begin{align} 
\delta (q)\;   \mathcal{K}^{(n)}_{\gamma} (s_{1}, \cdots s_{n-1})  &=\f{1}{(2\pi)^{n}} \int db e^{-iqb}\int da_{1} \cdots da_{n-1} \; e^{-i \sum_{k=1}^{n-1} s_{k}a_{k}} \;  \mathcal{K}^{(n)}_{\gamma} (a_{1}, \cdots a_{n-1}) \nonumber \\
&=\f{n}{(2\pi)^{n}}\int d\omega_{1} \cdots  d\omega_{n} \;  e^{-iqb}  e^{-i \sum_{k=1}^{n-1} s_{k}a_{k}} \; \mathcal{K}^{(n)}_{\gamma} (\omega_{1}, \cdots \omega_{n}),
\end{align}
in the second line we used the relations (\ref{eq; relation}).

%

Now the integral is 
\begin{align} 
\delta (q)\;   \mathcal{K}^{(n)}_{\gamma} (s_{1}, \cdots s_{n-1})  &= \f{n}{(2\pi)^{n}}\int d\omega_{1} \cdots  d\omega_{n}\; e^{-iqb}  e^{-i \sum_{k=1}^{n-1} s_{k}a_{k}}\; \mathcal{K}^{(n)}_{\gamma}(\omega_{1}, \cdots \omega_{n})  \nonumber \\
&=\f{n}{(2\pi)^{n}} \int_{C} \f{dz}{2\pi i} z^{\gamma}\int d \omega_{1}\f{e^{-\omega_{1}\left[ -2\pi(\gamma-1)+i(s_{1}+q)\right] }}{(z-e^{-2\pi \omega_{1}})^2} \nonumber \\
&\times \prod^{n-1}_{i=2} \int d \omega_{i} \f{e^{-\omega_{i}\left[ 2\pi +(s_{k}-s_{k-1}+q) i\right] }}{z-e^{-2\pi \omega_{i}}}\nonumber \\
&\times \int d \omega_{n} \f{e^{-\omega_{n}\left[ 2\pi -(s_{n-1}-q) i\right] }}{z-e^{-2\pi \omega_{n}}} \nonumber \\
& \equiv \f{n}{(2\pi)^{n}}  \int_{C} \f{dz}{2\pi i} J(z).  \label{eq;kerintefuza}
\end{align}

The strategy to compute this complicated integral is first compute each $\omega_{i}$ integral, and express 
$J(z)$ as a function of  modular times $\{{s_{i}}\}_{i=1 \cdots n-1}$. We then perform the $z$ integral 
by choosing the contour along the real axis, 
\be 
\delta (q)\;   \mathcal{K}^{(n)}_{\gamma} (s_{1}, \cdots s_{n-1})  =  \f{n}{(2\pi)^{n}} \int^{\infty} _{0} \f{d\beta}{2\pi i} \left(J(\beta-i\epsilon) -J(\beta +i\epsilon) \right), \quad \epsilon \rightarrow 0_{+} .
\ee

The details of the calculation can be found in Appendix (\ref{sec;kscalculation}) and here we only present the
final result for the kernel $ \mathcal{K}^{(n)}_{\gamma} (s_{1}, \cdots s_{n-1})$ , 

\be
 \mathcal{K}^{(n)}_{\gamma} (s_{1}, \cdots s_{n-1}) =\f{i}{8\pi^2} \left( \f{-i}{4\pi}\right)^{n-2}\f{ (s_{1}+2\pi i\gamma) \sin \pi \gamma}{\sinh\left(\f{s_{1}+2\pi i\gamma}{2} \right) \prod^{n-1}_{k=2} \sinh \left( \f{s_{k}-s_{k-1}}{2}\right) \sinh\left( \f{s_{n-1}}{2}\right)  } \label{eq:generala}
\ee

\vspace{1cm}
%
%
%
%
%
%
%

\subsection{ Choice of the integration contour: The quadratic $n=2$ term}  
In the previous subsection we derived the expression (\ref{eq:generala}) of the real time kernel $\mathcal{K}^{(n)}_{\gamma}(s_{1}, \cdots s_{n-1})$.
  In order to complete the discussion we need to properly fix the contour of the real time integrals $C$ in (\ref{eq;pertfinal}). We can do so by demanding  the Fourier transformation can be correctly reversed,
\be 
\mathcal{K}^{(n)}_{\gamma} (\omega_{1}, \cdots, \omega_{n}) = \int_{C_{k}} \prod_{k=1}^{n-1} ds_{k}\;  e^{i \sum_{k=1}^{n-1} s_{k}a_{k}} \; \mathcal{K}^{(n)}_{\gamma}(s_{1}, \cdots s_{n-1}). \label{eq;omegadef}
\ee

We first  consider the contour of  quadratic   $n=2$ term, 
\be
\int_{C_{s}} ds \; \mathcal{K}^{(2)}_{\gamma} ( s) \; e^{ias} = \f{i \sin \pi \gamma}{8 \pi^2}\int_{C_{s}} ds \; \f{s+2\pi i\gamma}{\sinh\f{s}{2} \sinh \f{s+2\pi i \gamma}{2}} \;e^{ias}, \label{eq:nontint2}
\ee
which is a bit tricky compared to higher order terms.   
When $a>0$ we  close the contour on the upper half plane.

The real time kernel  $\mathcal{K}^{(2)}_{\gamma} ( s)$ has two types of poles. 
\be 
s^{n}_{1}=2\pi i n,   \quad  s^{k}_{2}=2\pi i (k-\gamma), \quad n,k \in \mathbb{Z}, \quad k\neq 0.
\ee

\begin{figure}
\begin{center}
\includegraphics[width=8cm]{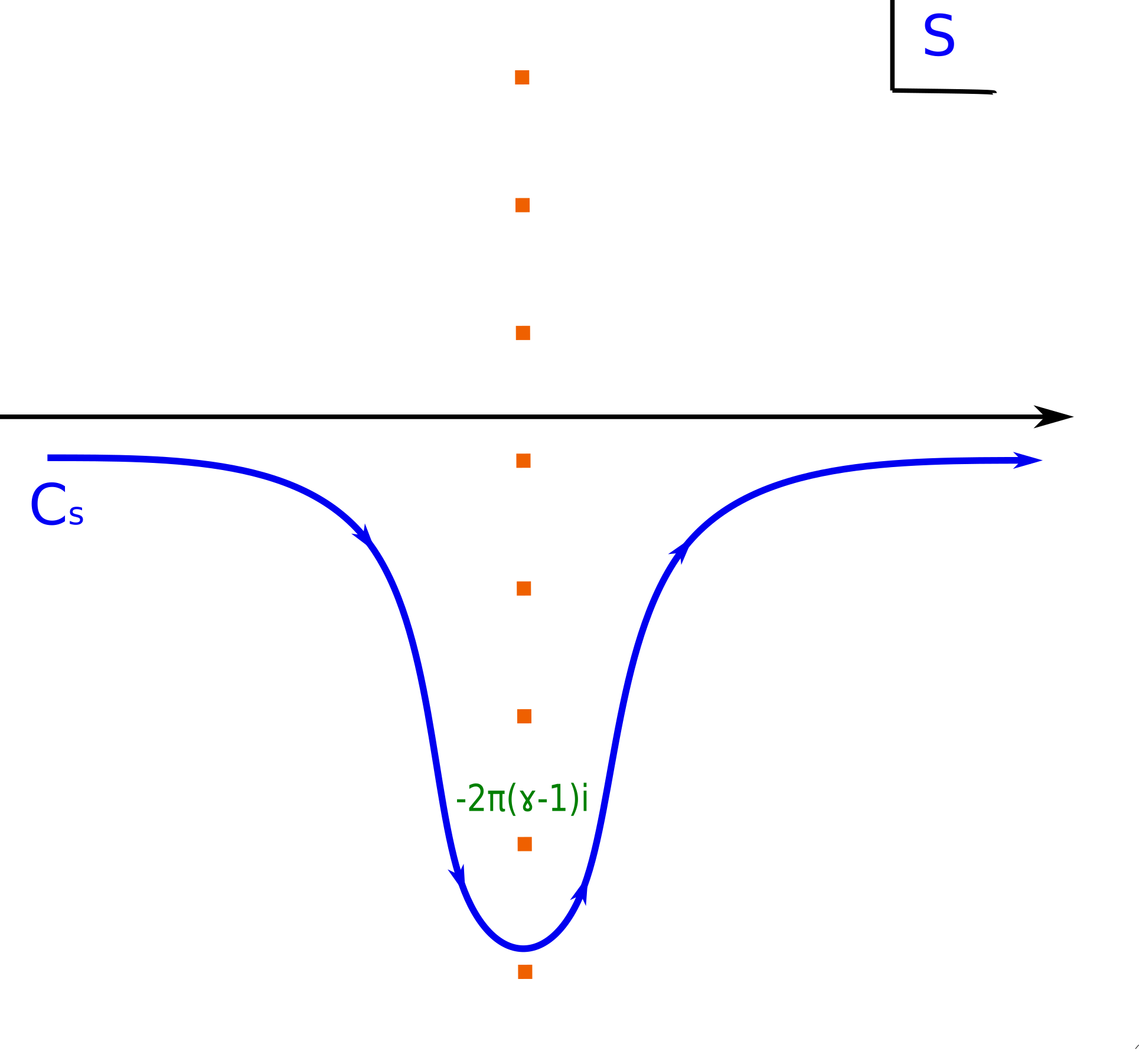}
\end{center}
\caption{The contour  $C_{s}$ of the integral (\ref{eq:nontint2})(blue line). Orange dots are the poles $s^{k}_{2}=2\pi i (k-\gamma) $ of the kernel  $\mathcal{K}^{(2)}_{\gamma} ( s) $. }
\label{fig;contour}
\end{figure}

We can easily see that if one choose the contour $C_{s}$ which contains $s^{n}_{1}, n\geq 0$, and $   s^{k}_{2}, k\geq 1$ (as in  figure \ref{fig;contour}), then the Fourier transformation is correctly reversed, 
\be
\mathcal{K}^{(n)}_{\gamma} (a)=\int_{C_{s}} ds \; \mathcal{K}^{(2)}_{\gamma} ( s) \; e^{ias}.
\ee

Again we explicitly check this in Appendix \ref{sec;inversen2}.

It is useful to write the integral as follows.
Since  we can write the integrand, 
\be 
\f{(s+2\pi i \gamma)\sin \pi \gamma}{\sinh \f{s}{2}\sinh \f{s+2\pi i \gamma}{2}} =\f{s+2\pi i \gamma}{1-e^{-s}}-\f{s+2\pi i \gamma}{1-e^{-(s+2\pi i \gamma)}}
\ee

then, the contour integral is naturally split into two parts, 
\be 
\int_{C_{s}} ds \f{(s+2\pi i \gamma)\sin \pi \gamma}{\sinh \f{s}{2}\sinh \f{s+2\pi i \gamma}{2}} G(s)= \int^{\infty-i \epsilon}_{-\infty-i \epsilon}ds \left[ \f{s+2\pi i \gamma}{1-e^{-s}} \right] G(s) -\int^{\infty-2\pi i (\gamma-\epsilon)}_{-\infty-2\pi i  (\gamma-\epsilon)} ds \left[\f{s+2\pi i \gamma}{1-e^{-(s+2\pi i \gamma)}} \right] G(s)
\ee

for any function $G(s)$ which is holomorphic  on the strip $ -2\pi \gamma < {\rm Im} s <0, $ when $\gamma>0$ .

It is also worth emphasizing that when $ -1<\gamma<1$, the contour gets simplified,  
\be 
\int_{C_{s}} ds \; \mathcal{K}^{(2)}_{\gamma} ( s) \; G(s)= \int^{\infty-i \epsilon}_{-\infty-i \epsilon} ds \; \mathcal{K}^{(2)}_{\gamma} ( s) \;  G(s)
\ee

\subsection{ Contour choice: $n \geq 3$ terms }

Now we fix all  the contours $C_{k} $ in the integral (\ref{eq;omegadef}).

In the above derivation we have used following formula, 
\be 
I_{1} (\xi, \beta+i\epsilon)=\int^{\infty}_{-\infty} d \omega \f{e^{-\omega \xi }}{(\beta+i\epsilon) -e^{-2\pi \omega}}  =\beta^{\left(\f{\xi}{2\pi}-1 \right)} \left( \f{e^{i\f{\xi}{2}}}{2\sin \f{\xi}{2}}\right).
\ee
Notice that $\xi=p+it$, and $p$ was a real number. In order for the integral to have an inverse, 
we need to make sure the choice of the contour $C_{t}$
\be 
\int_{C_{t}} dt \; I(p+it, \beta) e^{i\omega t} =\f{e^{-\omega p}}{\beta-e^{-2\pi \omega} }.
\ee
The integrand has poles at $s_{n}=ip +2\pi n$.  
By an explicit calculation, we recognize that we need pick up poles with $n\geq1$, thus 

\be 
\int_{C} dt  \equiv \int^{\infty +i(p+\epsilon)}_{-\infty+ i(p+\epsilon)} dt.
\ee

This in particular means that

\begin{align} 
\mathcal{K}_{n}^{\gamma}(\omega_{1}, \cdots \omega_{n}) 
&=\int_{C} \f{dz}{2\pi i} z^{\gamma}  \f{e^{2\pi (\gamma-1)\omega_{1}}}{(z-e^{-2\pi \omega_{1}})^{2}}\prod^{n}_{i=2} \f{e^{-2\pi \omega_{i}}}{z-e^{-2\pi \omega_{i}}} \nonumber \\
&= \prod^{n}_{k=1} \int^{\infty +i(p_{k}+\epsilon)}_{-\infty+ i(p_{k}+\epsilon)} dt_{k}e^{i \omega_{k} t_{k}}  \int_{C} \f{dz}{2\pi i}   z^{\gamma} \prod^{n}_{k=1}  I(it_{k}+p_{k}, z) \nonumber \\
&=\prod^{n-1}_{k=1} \int_{C_{k}} ds_{k} \;  e^{i\sum_{k=1}^{n-1} s_{k}a_{k}} \mathcal{K}_{n}^{\gamma} (s_{1}, \cdots s_{n-1})
\end{align}

Therefore  we need to choose the following contours, 
\be 
{\rm Im} s_{1}=-2\pi(\gamma-\epsilon), \quad {\rm Im} s_{k} - {\rm Im} s_{k-1} =\epsilon , \quad  {\rm Im} s_{n-1} =-\epsilon \label{eq;contourchoice}
\ee

 In particular when $\gamma<0$ there is no consistent contour choice for $n \geq 3$ terms.

\section{Applications to conformal field theory} 
\label{sec;cft}

The discussion so far is quite general, applicable to any density matrices  of any theories. From now on, 
we would like to apply the formula to  a special type  of reduced density matrices in conformal field theory(CFT). For this purpose, we first briefly summarize the construction of the reduced density matrices. For detailed discussions we refer to \cite{Sarosi:2017rsq}.

\subsection{Set up}

We start from a  CFT on  $d $ dimensional cylinder $\mathbb{R} \times S^{d-1}$, 
\be
ds^2=dt^2+ d\theta^{2} +\sin^2 \theta d\Omega_{d-2}^2. 
\ee 
 
We consider a ball shaped subsystem $A$, which is given by 
\be 
A: [0, \theta_{0}] \times S^{d-2}, \quad t=0, 
\ee 

 and a reduced density matrix $\rho_{V}$ of a globally  excited state $|V\ra$ on the region $A$ , 
\be
\rho_{V} = {\rm tr}_{A^{c}} |V \ra \la V |.  
\ee
The reduced density matrix has a path integral representation on the cylinder with a branch cut on A. The branched cylinder is mapped to $S^{1} \times H^{d-1}$ with the metric \cite{Casini:2011kv},
\be 
ds^{2} =d\tau^2+ du^2 + \sinh^2u d\Omega_{d-2}^2, \quad\tau \sim \tau +2\pi.  \label{eq;metric}
\ee

We find that  in this frame  $\rho_{V}$ has following expression \cite{Sarosi:2017rsq}, 
\be 
\rho_{V}=\f{ e^{-\pi K}V(\theta_{0}) V(-\theta_{0}) e^{-\pi K}}{\la V(\theta_{0}) V(-\theta_{0}) \ra}
\ee
where $K$ is the generator of the translation along $\tau$ direction, which can be identified with the modular Hamiltonian of $\rho_{0}$ and $V(\pm \theta_{0})$ are local 
operators corresponding to the excited states through state operator correspondence, located at 
$\tau = \pm \theta_{0}, u=0$.   In the small subsystem limit $\theta_{0} \rightarrow 0$, $V( \theta_{0}) \rightarrow V( -\theta_{0})$.

In this limit we can split the density matrix into the vacuum one  $ \rho_{0} =e^{-2\pi K}$ and the rest, $\rho_{V}=\rho_{0}+\delta \rho$. We do so by taking operator product expansion (OPE) of the  two local operators, 
\be 
\rho_{V}= \rho_{0}+ e^{-\pi K} \left[ \sum_{\mathcal{O}:{\rm primaries}} C^{\mathcal{O}}_{VV} B_{\mathcal{O}} (\theta_{0}, -\theta_{0})  \right]e^{-\pi K}
\ee 
where the index $\mathcal{O}$ labels  non identity primaries , and  $C^{\mathcal{O}}_{VV}, \;$ $ B_{\mathcal{O}} (\theta_{0}, -\theta_{0})$ are the OPE coefficient and the OPE block of $\mathcal{O}$ respectively.  

\subsection{ The perturbative expression of $T_{\gamma}(\rho) $}  

Now we determine the perturbative expression of $T_{\gamma} (\delta \rho)$ in CFT from (\ref{eq;pertfinal}). We write, 

\begin{align} 
{\rm tr} \rho^{\gamma} = {\rm tr} \rho_{0}^{\gamma}  +\sum T^{(n)}_{\gamma} (\delta \rho) ,
\end{align} 
 and for convenience we reproduce the expression of  $T^{(n)}_{\gamma}$ explicitly. 
\be
T^{(n)}_{\gamma} (\delta \rho)= \int ds_{1} \cdots ds_{n-1}  \mathcal{K}^{(n)}_{\gamma} (s_{1}, \cdots s_{n-1}) {\rm tr} \left[ e^{-2\pi \gamma K}\prod^{n}_{k=1} e^{i Ks_{k}} \tilde{\delta} \rho e^{-i Ks_{k}} \right] . \label{eq;Tn}
\ee 

Since $ \tilde{\delta} \rho = e^{\pi K} \delta \rho e^{\pi K} $,  in our case  we have 
\be
e^{i Ks}\delta \tilde{\rho}e^{-i Ks}=   \sum_{\mathcal{O}:{\rm primaries}} C^{\mathcal{O}}_{VV} \; B_{\mathcal{O}} (is+ \theta_{0},  is-\theta_{0}) .
\ee

For our $\delta \rho$, the trace in (\ref{eq;Tn}) can be regarded as a correlation function of the OPE blocks on the covering space $\Sigma_{\gamma} = S^{1}_{\gamma} \times H^{d-1}$, with the metric (\ref{eq;metric}) but the periodicity   of the Euclidean time direction is changed $\tau  \sim \tau +2\pi \gamma$,

\be
\la \cdots \ra_{\Sigma_{\gamma}} \equiv \f{1}{Z_{\gamma}} {\rm tr} \left[ e^{-2\pi \gamma K} \cdots \right],
\ee

where $Z_{\gamma}$ is  the CFT partition function on this space.

Combining these we can write each term of $T_{\gamma} (\delta \rho)$ by an integral of correlation functions of OPE blocks on $\Sigma_{\gamma}$ along  modular flow of vacuum $\rho_{0}$. 
\be
\f{1}{Z_{\gamma}}T^{(n)}_{\gamma} (\delta \rho)= \sum_{\{ \mathcal{O}_{l}\}} \; \prod^{n}_{l=1} \; C^{\mathcal{O}_{l}}_{VV} \int ds_{1} \cdots ds_{n-1} \mathcal{K}^{(n)}_{\gamma}  (s_{1}, \cdots s_{n-1}) \la  \prod^{n-1}_{k}B_{\mathcal{O}_{k}} (is_{k}+ \theta_{0},  is_{k}-\theta_{0})  B_{\mathcal{O}} (\theta_{0}, -\theta_{0})\ra_{\Sigma_{\gamma}}
\ee

\subsection{Bring $n=2$ term to the standard form} 
\label{subsec;sa}

We have seen $n=2$ term is given by 
\be 
\f{1}{Z_{\gamma}}T^{(2)}_{\gamma} (\delta \rho) = \sum_{\mathcal{O}: {\rm primaries}} \int_{C} ds \;  \mathcal{K}^{(2)}_{\gamma} (s) \la B_{\mathcal{O}} (is+ \theta_{0},  is-\theta_{0})  B_{\mathcal{O}} (\theta_{0}, -\theta_{0})\ra_{\Sigma_{\gamma}}
\ee

We can simplify this expression when  $0<\gamma <1$, and compare it with known results.  In order to do so, let us focus on the contribution $ T^{(2)}_{\gamma,\mathcal{O}} (\delta \rho)$ of a particular primary $\mathcal{O}$ to the $n=2$ term.  Since the OPE block $B_{\mathcal{O}}$ is summing up descendants of the primary $\mathcal{O}$,we can write it as
\be 
B_{\mathcal{O}}(\theta_{0}, -\theta_{0}) =C(\theta_{0}, \p_{a}) \mathcal{O}( \tau_{a}) \big|_{\tau_{a}=0}
\ee
where $C(\theta_{0}, \p_{a})$ is a differential  operator,  and $\tau_{a}$ is the coordinate of Euclidean timelike direction.  In the above we did not manifest the dependence of $\mathcal{O}$ on the coordinates of hyperbolic space.   
The  main  ingredient of the formula is the integral of two point function, 

\be 
I_{ab}=\f{i}{8\pi^2}\int^{\infty-i \epsilon}_{-\infty -i \epsilon} ds \f{s+2\pi i\gamma}{\sinh\f{s}{2} \sinh \f{s+2\pi i \gamma}{2}} G_{ab}(s) , \quad G_{ab}(s) =\la 
\mathcal{O}(is+\tau_{a}) \mathcal{O}(\tau_{b}) \ra_{\Sigma_{\gamma}} . \label{eq;IAB}
\ee 
and we can write,  
\be 
T^{(2)}_{\gamma,\mathcal{O}} (\delta \rho) =C(\theta_{0}, \p_{a}) C(\theta_{0}, \p_{b})  I_{ab}
\ee

As we explain in Appendix \ref{sec;n-2termsimp},  we can obtain a simpler expression of $T^{(2)}_{\gamma,\mathcal{O}} (\delta \rho) $,

\be
T^{(2)}_{\gamma,\mathcal{O}} (\delta \rho)= \f{\gamma\sin \pi \gamma}{4\pi} C(\theta_{0}, \p_{a})  C(\theta_{0}, \p_{b})\int^{\infty}_{-\infty}  \f{ds}{\sinh\f{s-\pi i\gamma}{2} \sinh \f{s+\pi i \gamma}{2}}  G_{ab}(s-\pi i\gamma) 
\ee

Notice that in the $\gamma \rightarrow 1$ limit,  its derivative   recovers the contribution of $\mathcal{O}$ to the second order term $S^{(2)}(\delta \rho )$ of  entanglement entropy \cite{Faulkner:2014jva},
\be 
S_{\mathcal{O}}^{(2)}(\delta \rho) = C(\theta_{0}, \p_{a}) C(\theta_{0}, \p_{b})  \int^{\infty }_{-\infty}ds  \f{-1}{4\sinh^2 \left( \f{s-i \epsilon}{2}\right)} \la \mathcal{O} (is +\tau_{a}) \mathcal{O} (\tau_{b})\ra_{\Sigma_{1}}.
\ee

\section{ Expansion of Petz's quasi entropy $D_{\gamma} (\rho|| \sigma)$ } 
\label{sec;holographic}

In this section, we consider a similar perturbative  expansion  for  Petz's  quasi entropy \cite{1985787},  defined by 

\be 
D_{\gamma} (\rho|| \sigma)= {\rm tr} \; \ \rho^{\gamma}\sigma^{1-\gamma} , \label{eq;relativerenyi}
\ee

In this section we  consider the case  where the one of the reduced density matrices is vacuum $\sigma= \rho_{0}$. We then write $\rho= \rho_{0} +\delta \rho$,
\be 
D_{\gamma} (\rho|| \rho_{0}) =\sum_{n=2}^{\infty} D^{(n)}_{\gamma} (\delta \rho).
\ee

The derivation of the perturbative series is very similar to the one  of $T_{\gamma}(\rho)$.  We first write
\be
D_{\gamma} (\rho|| \rho_{0})  = \int_{C}\f{dz}{2\pi i} \;  z^{\gamma}\;  {\rm tr}\; \f{\rho_{0}^{1-\gamma}}{z-\rho }, 
\ee

then by expanding the denominator we obtain a similar perturbative series.  One notable difference  is that in the power series of $D_{\gamma} (\rho|| \rho_{0})$, the 
$\rho^{\gamma}_{0}$ factor appears in the expansion (\ref{eq;Tn})  is canceled with the $\rho_{0}^{1-\gamma}$ factor which appear in the definition (\ref{eq;relativerenyi}).  The explicit expression of $D^{(n)}_{\gamma} (\delta \rho)$ is given by

\be
D^{(n)}_{\gamma} (\delta \rho)= \sum_{\{ \mathcal{O}_{k}\}} \; \prod^{n}_{k=1} \; C^{\mathcal{O}_{k}}_{VV}\prod^{n-1}_{k=1} ds_{k}\int ds_{k} \mathcal{K}^{(n)}_{\gamma}  (s_{1}, \cdots s_{n-1}) \la  \prod^{n-1}_{k}B_{\mathcal{O}_{k}} (is_{k}+ \theta_{0},  is_{k}-\theta_{0})  B_{\mathcal{O}_{n}} (\theta_{0}, -\theta_{0})\ra_{\Sigma_{1}} \label{eq; expansionD}
\ee
with the kernel $\mathcal{K}^{(n)}_{\gamma}  (s_{1}, \cdots s_{n-1})$ defined in  (\ref{eq:generala}).

One advantage of this quantity is that we can expand  it in terms of correlation functions on the space without branch cut, $\Sigma_{1}$, on the contrary to R\'enyi entropy itself, which is expanded  by correlators  $\la \cdots \ra_{\Sigma_{\gamma}}$ on the space $\Sigma_{\gamma}$ with branch cuts, and they are highly theory dependent quantities.  This implies that  first few terms of  $D_{\gamma} (\rho|| \sigma)$  are theory independent, and allows us to write them holographically.    

We also emphasize that the expressions (\ref{eq; expansionD}) are only valid in some range of $\gamma$. 
In particular higher order terms $D^{(n)}_{\gamma} (\delta \rho), \;  n \geq 3$,  has an expression in terms of a modular flow integral only in the range $0<\gamma <1$.
The limitation is again coming from the fact that there is a consistent contour choice of the modular flow integrals  (\ref{eq;contourchoice})  only in the range.  However $n=2$ term is still computable by the  modular flow integral for any value of $\gamma$.

Below, we will be focusing on following quantity, 
\be
Z_{\gamma}(\rho|| \sigma) \equiv  D_{-\gamma}(\rho|| \sigma) - D_{\gamma}(\rho|| \sigma) ,
\ee

and its quadratic part, 
\be 
Y_{\gamma}(\delta \rho) \equiv \f{d^{2}}{dt^{2}} Z_{\gamma}(\sigma + t\delta \rho\;  || \rho_{0})\big|_{t=0}, .
\ee

as well as its derivative with respect to the index $\gamma$,

\be 
X_{\gamma}(\delta \rho) = \f{d}{d \gamma} Y_{\gamma}(\delta \rho).
\ee

Notice that when $\gamma=0$  $\p_{\gamma} Z_{\gamma}(\rho|| \sigma) $ reduces to the relative entropy 
\be
\p_{\gamma} Z_{\gamma}(\rho|| \sigma) \big|_{\gamma=0} =     2S(\sigma|| \rho),
\ee
in which  the order of two density matrices is flipped $\rho \leftrightarrow \sigma$, and $X_{\gamma}(\delta \rho)$ reduces to the Fisher information, which is symmetric under the exchange.  
\be 
X_{\gamma}(\delta \rho)\big|_{\gamma=0} = F(\rho || \sigma).
\ee

\subsection{ Expressing $X_{\gamma}(\delta \rho)$ and $Y_{\gamma}(\delta \rho)$  by modular flow integrals}

Below we will focus on the range of the R\'enyi index  $-1 < \gamma <1$ for $D_{\gamma}^{(2)}( \delta \rho)$, or equivalently $0<\gamma<1$ for $X_{\gamma}(\delta \rho)$ and $Y_{\gamma}(\delta \rho)$ .

When the R\'enyi index is  in the window,  $Y_{\gamma}(\delta \rho)$ has following simple modular flow integral representation, 

\be 
Y_{\gamma}(\delta \rho) =\int^{\infty -i \epsilon}_{-\infty -i \epsilon}   \left[\mathcal{K}^{(2)}_{-\gamma}(s)-\mathcal{K}^{(2)}_{\gamma}(s) \right] {\rm tr}\left[e^{-2\pi K} \tilde{\delta} \rho (s) \; \tilde{\delta} \rho \right] ds.
\ee

$\mathcal{K}^{(2)}_{\gamma}(s) $ is given by 
\be 
\mathcal{K}^{(2)}_{\gamma}(s) =\f{i \sin \pi \gamma}{8\pi^2} \f{(s+2\pi i \gamma)  }{\sinh \f{s}{2} \sinh \f{s+2\pi i \gamma}{2}}.
\ee

For the class of $\delta \rho$ we are interested in,  we have
\begin{align}
Y_{\gamma}(\delta \rho)&= C(\theta_{0}, \p_{a})C(\theta_{0}, \p_{b}) \int^{\infty -i \epsilon}_{-\infty -i \epsilon} 
 \left[\mathcal{K}^{(2)}_{-\gamma}(s)-\mathcal{K}^{(2)}_{\gamma}(-s-2\pi i) \right] \la \mathcal{O} (is +\tau_{a}) \mathcal{O} (\tau_{b})\ra_{\Sigma_{1}} ds \nonumber \\[+10pt] 
&= C(\theta_{0}, \p_{a})C(\theta_{0}, \p_{b}) \int^{\infty }_{-\infty}ds  \left[ \f{-(\sin \pi \gamma) /4\pi }{\sinh\left( \f{s-i \epsilon}{2}\right)\sinh\left( \f{s-2\pi i \gamma}{2}\right)} \right]\la \mathcal{O} (is +\tau_{a}) \mathcal{O} (\tau_{b})\ra_{\Sigma_{1}} . \label{eq;ygamma}
\end{align}

In the second term of the first line, we used another expression of $D^{2}_{\gamma} (\delta \rho)$ 
\be 
D^{2}_{\gamma} (\delta \rho)= C(\theta_{0}, \p_{a})C(\theta_{0}, \p_{b}) I_{ba}, \quad I_{ba} =\int^{\infty+i \epsilon}_{-\infty+i \epsilon} ds \mathcal{K}^{(2)}_{\gamma}(s-2\pi i)  \la \mathcal{O} (is +\tau_{b}) \mathcal{O} (\tau_{a})\ra_{\Sigma_{1}}, \quad \tau_{a} > \tau_{b},
\ee
and flipped the sign of the integration variable $s \rightarrow -s$. The derivation of this expression is  the same with that of (\ref{eq;Ibaa}) in Appendix \ref{sec;n-2termsimp}.

By taking derivative of (\ref{eq;ygamma}) with respect to $\gamma$, we have an expression of $X_{\gamma}(\delta \rho)$, 
\be
X_{\gamma}(\delta \rho)=C(\theta_{0}, \p_{a})C(\theta_{0}, \p_{b}) \int^{\infty }_{-\infty}ds  \f{-1}{4\sinh^2 \left( \f{s-2\pi i \gamma}{2}\right)} \la \mathcal{O} (is +\tau_{a}) \mathcal{O} (\tau_{b})\ra_{\Sigma_{1}}.  \label{eq;xgamma}
\ee

\subsection{Holographic expressions  of $X_{\gamma}(\delta \rho)$ and $Y_{\gamma}(\delta \rho)$ } 

So far we have obtained quadratic term $Y_{\gamma}(\delta \rho)$   which is  particular linear combination of  the R\'enyi relative divergence $Z_{\gamma}(\delta|| \rho_{0})$, and its derivative  $X_{\gamma}(\delta \rho)$  in terms of modular flow integral (\ref{eq;ygamma}), (\ref{eq;xgamma}). 

%
%
 
As we will see below, through AdS/CFT correspondence, they  have  simple bulk expressions. The derivations are parallel to the argument of \cite{Faulkner:2017tkh}, where they obtained the holographic expression of quadratic term of the entanglement entropy $S^{(2)}(\delta \rho)$. 
\subsubsection{Set up}

 To explain this let us first recall the corresponding bulk set up. Our reference state is the vacuum reduced density matrix $\rho_{0}$, and since we take the subsystem $A$ to be a ball shape region, corresponding Ryu Takayanagi surface can be regarded as the bifurcation surface $r_{B}=1$ of the topological black hole, 
\be 
ds^2=-(r_{B}^2-1) ds_{B}^2 + \f{dr_{B}^2}{(r_{B}^2-1)} +r_{B}^2 dH_{d-1}^2 \label{eq;TBH}
\ee

where $dH_{d-1}^2$ denotes the metric of $d-1$ dimensional hyperbolic space, 
\be
dH_{d-1}^2 = du^2 + \sinh^2 u \; d\Omega_{d-2}^2.
\ee 

In \cite{Faulkner:2017tkh}  it was shown that the CFT two point function in  (\ref{eq;ygamma}), (\ref{eq;xgamma})  can be written in terms of the  bulk symplectic form $\omega_{\phi}$ of the bulk field $\phi$ dual to the CFT primary $\mathcal{O}$,  
\be 
\la \mathcal{O} (is +\tau_{a}) \mathcal{O} (\tau_{b})\ra_{\Sigma_{1}} =- \int dX_{B} \; {\boldmath \omega}_{\phi} \left(K_{E}(X_{B}| \tau_{ba}), K_{R}(X_{B}| s) \right)  \label{eq;twopt}. 
\ee

We evaluate the integral on fixed $r_{B}=r_{0}$ surface of the topological black hole (\ref{eq;TBH}), and collectively denote the coordinates of the surface by $X_{B}$.  The bulk symplectic form is given by 
\be
 \omega_{\phi} (\delta \phi_{1}, \delta \phi_{2}) = n^{M} \left( \delta\phi_{1} \p_{M} \delta\phi_{2} - \delta\phi_{2} \p_{M} \delta\phi_{1}  \right),
\ee
where $n^{M}$ is the normal vector of the $r_{B} =r_{0}$ surface. 
$K_{E}( X_{B}| \tau_{ba})$,   $K_{R}( X_{B}|s)$ are  the Euclidean and Retarded bulk to boundary propagator of the bulk field $\phi$, respectively. The primary operators in the CFT  two point function are located at the origin of the hyperbolic space $u=0$. We omit this information  in the bulk to boundary propagators.  

\subsubsection{Holographic rewritings}
\label{subsubsec;HR}

By plugging  (\ref{eq;twopt}) into (\ref{eq;ygamma}), and evaluating the remaining $s$ integral  by picking up poles of the kernel, we get \footnote{The argument here is very similar to the one in \cite{Faulkner:2017tkh}.  See Appendix \ref{sec;rewritings} for the details.}

\be 
Y_{\gamma}(\delta \rho)=i \; C(\theta_{0}, \p_{a})C(\theta_{0}, \p_{b}) \int dX_{B} \;  \omega_{\phi}\left( K_{E}(X_{B}| \tau_{ba}),\; K_{E} (X_{B} |-2\pi  \gamma)-  K_{E} (X_{B} |0) \right)
\ee

By shifting the time coordinate $s_{B} \rightarrow s_{B} +i \tau_{a}$, and using the relation between the Euclidean bulk to boundary propagator  and the expectation value of the bulk scalar field operator $ \phi (X_{B})$, 
\be 
 C(\theta_{0}, \p_{a}) K_{E} (X_{B}| \tau_{a}) = \la V | \phi (X_{B}) |V \ra \equiv \la \phi (X_{B}) \ra_{V},
\ee

we get, 
\be 
Y_{\gamma}(\delta \rho) =i \int dX_{B} \;  \omega_{\phi} \left(\la \phi (0) \ra_{V},\; \la \phi (2\pi  \gamma) \ra_{V} -\la \phi (0) \ra_{V} \right). 
\ee

where  $\la \phi (2\pi  \gamma) \ra_{V}$ is the expectation value of the  bulk field rotated by $2 \pi \gamma$ along the Euclidean timelike direction,

\be 
\la \phi (2\pi  \gamma)  \ra_{V} \equiv {\rm tr} \left[  \rho_{V}\;  e^{-2\pi\gamma K}\;\phi \; e^{2\pi \gamma K} \right]
\ee

In the argument of the bulk local field $\phi$, we only manifested the  Euclidean time like coordinate, 
\be
\phi (\tau)  \equiv \phi (r_{B}, \tau+is_{B}, u, \Omega_{d-2}) .
\ee

We can obtain a similar expression for $X_{\gamma}(\delta \rho)$ just by  taking a derivative of $Y_{\gamma}(\delta \rho)$, 
\be 
X_{\gamma}(\delta \rho)= -2\pi \int dX_{B} \;  \omega_{\phi}\left( \la \phi (0) \ra_{V},\;  \p_{s}\la \phi (2\pi  \gamma)  \ra_{V} \right),
\ee
here we used the relation $\p_{\gamma} =-i\p_{s}$.
This integral is invariant under the deformation of the surface on which we are evaluating the integral. In particular we can choose the fixed time slice $s_{B} =0$, then the integral can be written as, 
\be 
X_{\gamma}(\delta \rho)= -2\pi \int_{\Sigma} d\Sigma^{a} \; \xi^{b} \; T_{ab} (\la \phi (0) \ra_{V}, \; \la \phi (2\pi  \gamma)  \ra_{V}),
\ee

where $\Sigma$ is the bulk region  on the time slice  $s_{B} =0$, which is enclosed by the boundary subsystem A and the bifurcation surface of the topological black hole (ie,  RT surface).  Also $d\Sigma^{a}$ is the volume element of $\Sigma$, and $\xi^{b}$ is the timelike Killing vector of the black hole.   $T_{ab}$ is a quadratic form of $\phi$ related to  the stress energy tensor of the bulk field, 
\be 
T_{ab} (\phi_{1}, \phi_{2}) = \p_{a}\phi_{1} \p_{b}\phi_{2} - m^{2}g_{ab} \phi_{1} \phi_{2} 
\ee

There is another way to derive this result. Let us come back to the CFT formula, 
\be
X_{\gamma}(\delta \rho)=C(\theta_{0}, \p_{a})C(\theta_{0}, \p_{b}) \int^{\infty }_{-\infty}ds  \f{-1}{4\sinh^2 \left( \f{s-2\pi i \gamma}{2}\right)} \la \mathcal{O} (is +\tau_{a}) \mathcal{O} (\tau_{b})\ra_{\Sigma_{1}}
\ee 

by changing the integration variable to $t =s -2\pi i \gamma$ and shifting the contour we get, 
\begin{align} 
X_{\gamma}(\delta \rho)&=C(\theta_{0}, \p_{a})C(\theta_{0}, \p_{b}) \int^{\infty }_{-\infty}dt  \f{-1}{4\sinh^2 \left( \f{t-2\pi i \epsilon}{2}\right)} \la \mathcal{O} (i(t+2\pi i \gamma) +\tau_{a}) \mathcal{O} (\tau_{b})\ra_{\Sigma_{1}} \nonumber \\[+10 pt]
&=  \int^{\infty }_{-\infty}ds \f{-1}{4\sinh^2 \left( \f{s-2\pi i \epsilon}{2}\right)} {\rm tr} \left[ \tilde{\delta} \rho (s)\;  e^{2\pi \gamma}\; \tilde{\delta}\rho \; e^{-2\pi\gamma}  \right] \label{eq;deltaK}
\end{align} 

In \cite{Sarosi:2017rsq} it was shown that the excited state modular Hamiltonian $K_{\rho}$ of $\rho$, when expanded by $\delta \rho$,  the leading  order correction to the vacuum modular Hamiltonian $K$ is  given by 
\be 
K_{\rho} = K + \int^{\infty}_{-\infty} \f{ds}{\sinh^{2} \f{s}{2}} \tilde{\delta} \rho (s)  \equiv K+\delta K.
\ee

It was also shown that contribution of a primary operator $\mathcal{O}$ to the correction $\delta K$  has a bulk expression 
\be 
\delta K  =  2\pi \int_{\Sigma} d\Sigma^{a} \; \xi^{b} \; T_{ab} (\la \phi (0) \ra_{V}, \; \hat{\phi}),
\ee
where $\hat{\phi}$ is the bulk field operator dual to $\mathcal{O}$. By plugging this into (\ref{eq;deltaK}), we recover the result.

\section{Conclusions}
\label{sec;conc}

In this paper we developed a novel  way to perturbatively expand R\'enyi tpe quantities involving powers of 
reduced density matrices.  
We then obtained a holographic expression of  the quadratic parts of R\'enyi relative divergences $X_{\gamma} (\delta \rho), Y_{\gamma} (\delta \rho)$ in terms of bulk symplectic form starting from the CFT calculations. 

\vspace{0.1cm}

It is interesting find  a bulk derivation of this result. One difficulty in doing so is coming form the fact that in general there is no nice path integral  representation of 
 R\'enyi relative divergence. This is because even if reduced density matrices $\rho, \sigma$ can be written by path integrals,  $\rho^{\gamma}$ and   $\sigma^{1-\gamma}$ can not. 
 If we could find such a representation, then we can map the CFT  path integral calculationss to the bulk on shell action calculations.
Indeed, in a special case where R\'enyi relative divergence can be represented by a path integral,  corresponding holographic calcuation is known\cite{Bernamonti:2018vmw}.   However in order to derive a bulk formula for 
 R\'enyi relative divergence between two generic bulk configurations, we need to take a different approach. 
A possible approach would be  first going back to replica trick \cite{Lewkowycz:2013nqa}, compute ${\rm tr} \rho^{n} \sigma^{m} $ for positive integers $n,m$ then analytically continue the result $n \rightarrow \gamma, m\rightarrow 1-\gamma$.  

\vspace{0.1cm}

Furthermore it would be nice if we could read off finer information of bulk geometries using  R\'enyi relative divergence.  It has been shown that using relative entropy, we can read off  first non linear part 
of Einstein equations \cite{Faulkner:2014jva,Faulkner:2017tkh} in particular.  Since  R\'enyi relative divergence is a one parameter generalization of relative entropy, and knows about details of eigenvalue distribution of excited state reduced density matrices, it is natural to expect this. 

\vspace{0.1cm}

 Another interesting direction would be   to calculate correlation functions with insertions of modular flows of excited states, by using the technique developed in this paper. For example\cite{Faulkner:2018faa,Faulkner:2017vdd,Chen:2018rgz}, two point function with an  insertion of a modular flow  $\la \mathcal{O}(x) \Delta^{it}  \mathcal{O}(y) \ra$ was considered. There, it was  also argued that this is useful to extract information of corresponding bulk geometry. 
  Naively speaking we can perturbatively compute them by Wick rotating the R\'enyi index $ \gamma$ to the imaginary value  $ \gamma \rightarrow it$ in our result.  The task would be to check  that there is no obstacle to do  this.

\section*{Acknowledgments}

We thank Alex Belin, Tom Faulkner, Sudip Ghosh, Norihiro Iizuka, Robert Myers,Tatsuma Nishioka, Jonathan  Oppenheim, G\'abor S\'arosi,  Tadashi Takayanagi  and Kotaro Tamaoka for discussions.

\appendix 
\section{The calculation of $\mathcal{K}^{(n)}_{\gamma} (s_{1}, \cdots s_{n-1})$}
\label{sec;kscalculation} 

In this appendix, we explain the details of the calculation of the kernel $\mathcal{K}^{(n)}_{\gamma} (s_{1}, \cdots s_{n-1})$, starting from  (\ref{eq;kerintefuza}).

 In order to do  this, we first decompose  $J(z)$ in (\ref{eq;kerintefuza})
\be 
J(z)  = z^{\gamma} I_{2}(\xi_{1}, z ) \prod^{n-1}_{k=2} I_{1}(\xi_{k}, z )   I_{1}(\xi_{n}, z ) ,
\ee

where  

\be
\xi_{1}=-2\pi(\gamma-1)+i(s_{1}+q), \quad \xi_{n}= 2\pi -(s_{n-1} -q) i, \label{eq;xi1}
\ee

\be
\xi_{k}=2\pi +(s_{k}-s_{k-1}+q)i, \quad 2 \leq k \leq n-1.\label{eq;xik}
\ee

and 
\begin{align} 
I_{1} (\xi, z)&=\int^{\infty}_{-\infty} d \omega \f{e^{-\omega \xi }}{z-e^{-2\pi \omega}}, \qquad I_{2} (\xi, z)=\int^{\infty}_{-\infty} d \omega \f{e^{-\omega \xi }}{(z-e^{-2\pi \omega} )^2}   .
\end{align}

For $I_{1} (\xi, z)$, by carefully picking up the contributions of the relevant poles we have, 
\be 
I_{1} (\xi, \beta+i \epsilon) =\beta^{\left(\f{\xi}{2\pi}-1 \right)} \left( \f{e^{-i\f{\xi}{2}}}{2\sin \f{\xi}{2}}\right), \quad  I_{1} (\xi, \beta-i \epsilon) =\beta^{\left(\f{\xi}{2\pi}-1\right)} \left( \f{e^{i\f{\xi}{2}}}{2\sin \f{\xi}{2}}\right) .\label{eq:discint}
\ee

One way to check these is using 
\be 
\f{1}{z+i \epsilon}- \f{1}{z-i \epsilon} =-2\pi i \delta(z),
\ee

Then,

\begin{align} 
{\rm Disc} \; I &= \lim_{\epsilon \rightarrow 0_{+}} \left[ I(z+i \epsilon) -I(z-i \epsilon) \right] \nonumber \\
&=-2\pi i  \int^{\infty}_{-\infty} d\omega e^{-\xi \omega}\; \delta(\beta- e^{-2\pi \omega}) =-i\beta^{\left(\f{\xi}{2\pi}-1 \right)}.
\end{align}

This is consistent with (\ref{eq:discint}).

We can  evaluate   $I_{2} (\xi, z)$ just by taking  derivative of  $I_{1} (\xi, z)$ with respect to $\beta$, 
\be 
I_{2} (\xi, \beta+i \epsilon)=-\left(\f{\xi}{2\pi}-1 \right) \beta^{\left(\f{\xi}{2\pi}-2 \right)} \left( \f{e^{-i\f{\xi}{2}}}{2\sin \f{\xi}{2}}\right),  \quad I_{2} (\xi, \beta-i \epsilon)=-\left(\f{\xi}{2\pi}-1 \right)\beta^{\left(\f{\xi}{2\pi}-2 \right)} \left( \f{e^{i\f{\xi}{2}}}{2\sin \f{\xi}{2}}\right).
\ee

Combining these, we obtain the relevant  expressions of $J(z)$
\be
J(\beta+i \epsilon) =-\beta^{\left(\gamma +\sum^{n}_{k=1} \f{\xi_{k}}{2\pi}-(n+1) \right)}  \f{\left(\f{\xi_{1}}{2\pi}-1\right) }{\prod^{n}_{k=1} 2\sin \f{\xi_{k}}{2}} e^{-\f{i}{2}\sum^{n}_{k=1}\xi_{k}},
\ee

and 
\be
J(\beta-i \epsilon) = -\beta ^{\left(\gamma +\sum^{n}_{k=1} \f{\xi_{k}}{2\pi}-(n+1) \right)}  \f{\left(\f{\xi_{1}}{2\pi}-1\right) }{\prod^{n}_{k=1} 2\sin \f{\xi_{k}}{2}} e^{\f{i}{2}\sum^{n}_{k=1}\xi_{k}}.
\ee

Since 
\be
\gamma + \sum^{n}_{k=1} \f{\xi_{k}}{2\pi}  -(n+1)= -1 + \f{iqn}{2\pi},
\ee

the $\beta$ integral  produces the delta function,  
\begin{align}
 \int^{\infty}_{-\infty} \f{d\beta}{2\pi i} \; \beta^{-1+\f{iqn}{2\pi} }=\f{2\pi }{ni} \delta(q).
\end{align} 

By picking up the discontinuity across the real line, we get

\begin{align}
\delta (q)\;   \mathcal{K}^{(n)}_{\gamma} (s_{1}, \cdots s_{n-1})  &= \f{n}{(2\pi)^{n}}\int^{\infty} _{0} \f{d\beta}{2\pi i} \left(J(\beta-i\epsilon) -J(\beta +i\epsilon) \right), \quad \epsilon \rightarrow 0_{+}, \nonumber \\
&= - \f{n}{(2\pi)^{n}} \left(\f{2\pi }{ni} \delta(q) \right) \f{\left(\f{\xi_{1}}{2\pi}-1\right) }{\prod^{n}_{k=1} 2\sin \f{\xi_{k}}{2}}\left(  e^{\f{i}{2}\sum^{n}_{k=1}\xi_{k}} - e^{-\f{i}{2}\sum^{n}_{k=1}\xi_{k}} \right)
\end{align}

Notice that 
\begin{align} 
 e^{-\f{i}{2}\sum^{n}_{k=1}\xi_{k}} - e^{+\f{i}{2}\sum^{n}_{k=1}\xi_{k}} &=e^{i\pi(\gamma-n)} -e^{-i\pi(\gamma-n)} \\
&=2i(-1)^{n} \sin \pi \gamma
\end{align}

and 
\be
\f{\left(\f{\xi_{1}}{2\pi}-1\right) }{\prod^{n}_{k=1} \sin \f{\xi_{k}}{2}}= \f{-i^{n+1}}{2\pi} \f{(s_{1}+2\pi i\gamma)}{\sinh\left(\f{s_{1}+2\pi i\gamma}{2} \right) \prod^{n-1}_{k=2} \sinh \left( \f{s_{k}-s_{k-1}}{2}\right) \sinh\left( \f{s_{n-1}}{2}\right)  }
\ee

\vspace{1cm}

From this   we  finally arrive at the expression of the kernel,
\be
 \mathcal{K}^{(n)}_{\gamma} (s_{1}, \cdots s_{n-1}) =\f{i}{8\pi^2} \left( \f{-i}{4\pi}\right)^{n-2}\f{ (s_{1}+2\pi i\gamma) \sin \pi \gamma}{\sinh\left(\f{s_{1}+2\pi i\gamma}{2} \right) \prod^{n-1}_{k=2} \sinh \left( \f{s_{k}-s_{k-1}}{2}\right) \sinh\left( \f{s_{n-1}}{2}\right)  } \label{eq:general}
\ee 

\section{Fixing the contour of $n=2$ term}
\label{sec;inversen2}

In this appendix, we fix the correct contour $C_{s}$ of $ n=2$ real time integral 

\be
\int_{C_{s}} ds \; \mathcal{K}^{(2)}_{\gamma} ( s) \; e^{ias} = \f{i \sin \pi \gamma}{8 \pi^2}\int_{C_{s}} ds \; \f{s+2\pi i\gamma}{\sinh\f{s}{2} \sinh \f{s+2\pi i \gamma}{2}} \;e^{ias}, \label{eq:nontint}
\ee

which reproduces the kernel in the frequency representation,
 (\ref{eq:kfreaq})

\begin{align}
\mathcal{K}^{(2)}_{\gamma}(\omega_{1}, \omega_{2})&=e^{2\pi \gamma \omega_{1} }e^{-2\pi \omega_{1}-2\pi \omega_{2} } K(\omega_{1},\omega_{2}) \nonumber \\ 
&=\f{e^{2\pi \gamma \omega_{1} }e^{-2\pi \omega_{1}-2\pi \omega_{2} } }{(e^{-2\pi \omega_{1}}-e^{-2\pi \omega_{2}})^{2}} \left[ (\gamma-1) e^{-2\pi \gamma \omega_{1}} +e^{-2\pi \gamma \omega_{2}} -\gamma  e^{-2\pi (\gamma-1) \omega_{1}}e^{-2\pi \omega_{2}}\right].
\end{align}

Using $a\equiv \omega_{1}-\omega_{2}$, we have, 
\be 
\mathcal{K}^{(2)}_{\gamma}(a)= \f{e^{2\pi a}}{(1-e^{2\pi a})^2}\left[ (\gamma-1) +e^{2\pi a\gamma} -\gamma e^{2\pi a}  \right].
\ee

Let's do the integral (\ref{eq:nontint}). There are two types of poles. 
\be 
s^{n}_{1}=2\pi i n,   \quad  s^{k}_{2}=2\pi i (k-\gamma), 
\ee

We choose a contour which contains $s^{n}_{1}, n\geq 0$, and $   s^{k}_{2}, k\geq 1$.  One way to manifest the 
contour prescription is introducing an additional parameter $x>0$, 
\be 
\mathcal{K}^{(2)}_{\gamma}(s,x) = \left(\f{i \sin \pi \gamma}{8 \pi^2} \right)\f{s+2\pi i\gamma}{\sinh\f{s+x}{2} \sinh \f{s+2\pi i \gamma}{2}},
\ee
and finally send $x \rightarrow 0$ to get the desired result.  


We have 
\be 
{\rm Res}[s^{1}_{n} ]=\f{ i(n+\gamma)}{2\pi  }e^{-2\pi an }, \quad {\rm Res}[s^{2}_{n} ] =-\f{ i k}{2\pi }e^{2\pi a(\gamma-k) }.
\ee

By combining them, 
\begin{align} 
\int_{C_{s}} ds \; \mathcal{K}^{(2)}_{\gamma} ( s) \; e^{ias} &=
2\pi i \left(\sum_{n}{\rm Res}[s^{1}_{n} ] +\sum_{k}{\rm Res}[s^{2}_{k} ] \right) \nonumber \\
& =-\left[ (1-e^{-2\pi a\gamma }  )\sum_{k} k e^{-2\pi ak } +\gamma \sum_{n} e^{-2\pi a\gamma n }   \right] \nonumber \\
&= -\f{e^{2\pi a }}{(e^{2\pi a  }-1)^{2}} \left (1-\gamma) + \gamma  e^{2\pi a  } -e^{2\pi a\gamma  } \right]\nonumber \\[+10pt]
&=  \mathcal{K}^{(2)}_{\gamma}(a).
\end{align}

This is what we want. In the sum, we included $n=0$ contribution. 

%
%
%

\section{Simplifying $T^{(2)}_{\gamma} (\delta \rho) $}
\label{sec;n-2termsimp}

In this section we simplify  $n=2$ term of $T^{(2)}_{\gamma} (\delta \rho) $. In section \ref{subsec;sa} we saw that the contribution of  particular primary $\mathcal{O}$ to $T^{(2)}_{\gamma} (\delta \rho) $ can be written 

\be 
T^{(2)}_{\gamma,\mathcal{O}} (\delta \rho) =C(\theta_{0}, \p_{a}) C(\theta_{0}, \p_{b})  I_{ab}
\ee

where
\be 
I_{ab}=\f{i}{8\pi^2}\int^{\infty-i \epsilon}_{-\infty -i \epsilon} ds \f{s+2\pi i\gamma}{\sinh\f{s}{2} \sinh \f{s+2\pi i \gamma}{2}} G_{ab}(s) , \quad G_{ab}(s) =\la 
\mathcal{O}(is+\tau_{a}) \mathcal{O}(\tau_{b}) \ra_{\Sigma_{\gamma}} . \label{eq;IAB}
\ee 

and  $C(\theta_{0}, \p_{a}) $ is a differential operator summing up all descendants.

 This expression  
only holds  when $\tau_{a}>\tau_{b}$ . This is because  we started from the spectral representation, 

\be
I_{ab}=  \int  d\omega_{1} d\omega_{2} \; \mathcal{K}^{\gamma} (a) \; e^{-2\pi \gamma \omega_{1}}  \la \omega_{1}  | \mathcal{O}(\tau_{a}) |\omega_{2}  \ra \la \omega_{2}  | \mathcal{O}(\tau_{b}) |\omega_{1}  \ra,
\ee 

rewrote  it  in terms of  the modular flow integral by 
\be 
\mathcal{K}^{\gamma} (a)  =\int^{\infty -i\epsilon}_{-\infty -i\epsilon} \; \mathcal{K}^{\gamma}(s)  e^{ias}, \quad a= \omega_{1}- \omega_{2}.
\ee

and then undoing the spectral decomposition of the two point function  $G_{ab}(s) $, 
\be
 \int  d\omega_{1} d\omega_{2}\; e^{-2\pi \gamma \omega_{1}+ias}  \la \omega_{1}  | \mathcal{O}(\tau_{a}) |\omega_{2}  \ra \la \omega_{2}  | \mathcal{O}(\tau_{b}) |\omega_{1}  \ra = \la 
\mathcal{O}(is+\tau_{a}) \mathcal{O}(\tau_{b}) \ra_{\Sigma_{\gamma}}.
\ee
The spectral integral only converges when  $\tau_{a}>\tau_{b}$.

When $\tau_{b} > \tau_{a}$, we instead write 
\be 
I_{ab}=\int  d\omega_{1} d\omega_{2} \; \left[\mathcal{K}^{\gamma} (a) e^{-2\pi \gamma a} \right]\; e^{-2\pi \gamma \omega_{2}}    \la \omega_{1}  | \mathcal{O}(\tau_{a}) |\omega_{2}  \ra \la \omega_{2}  | \mathcal{O}(\tau_{b}) |\omega_{1}  \ra, \label{eq:plus} 
\ee

\begin{align} 
\mathcal{K}^{\gamma} (a) e^{-2\pi \gamma a}  &=\int^{\infty -i\epsilon}_{-\infty -i\epsilon}  ds \; \mathcal{K}^{\gamma}(s) \; e^{ia(s+2\pi i \gamma) } \nonumber \\
&=\int^{\infty+2\pi i(\gamma -\epsilon)}_{-\infty+2\pi i(\gamma -\epsilon)} dt \;  \mathcal{K}^{\gamma}(t-2\pi i \gamma ) \;  e^{iat}.
\end{align} 
Since 
\be 
\mathcal{K}^{\gamma}(t-2\pi i \gamma ) =\f{i \sin \pi \gamma}{8\pi^2}\f{t}{\sinh \f{t}{2} \sinh \f{t-2\pi i\gamma}{2}}
\ee

is regular on the strip $ 2\pi (\gamma-\epsilon)>{\rm Im} t > 0$ when $0<\gamma<1$, we deform the contour to ${\rm Im} \; t  = \epsilon $
\be 
\mathcal{K}^{\gamma} (a) e^{-2\pi \gamma a} =\int^{\infty +i\epsilon}_{-\infty +i\epsilon} \; dt\; \mathcal{K}^{\gamma}(t-2\pi i \gamma)  \;  e^{iat}
\ee

Therefore for  $\tau_{b} > \tau_{a}$ we have 
\be 
I_{ab} =\int^{\infty +i\epsilon}_{-\infty +i\epsilon}\mathcal{K}^{\gamma}(s-2\pi i \gamma) \la \mathcal{O}(\tau_{b})   \mathcal{O}(\tau_{a} +is) \ra_{\Sigma_{\gamma}},  \quad \tau_{b} >\tau_{a} \label{eq:minus} .
\ee

We have  similar formule for $I_{ba}$, just by flipping  $\tau_{a} \leftrightarrow \tau_{b}$.

Finally we combine these expressions to get a simpler form of $T^{(2)}_{\gamma} (\delta \rho)$.  The two point  function in (\ref{eq;IAB}) is analytic in the strip region $-2\pi \gamma < {\rm Im} s < \tau_{ba}$.  Since when 
 $0< \gamma <1$ there is no pole coming from the kernel in the strip, and we are allowed to deform the contour $s \rightarrow s-\pi i\gamma $. Then the integral for $\tau_{a} > \tau_{b}$ becomes 
\be 
I_{ab}=\f{i\sin \pi \gamma}{8\pi^2}\int^{\infty-i \epsilon}_{-\infty -i \epsilon} ds \f{s+\pi i\gamma}{\sinh\f{s-\pi i\gamma}{2} \sinh \f{s+\pi i \gamma}{2}} G_{ab}(s-\pi i\gamma), \quad \tau_{a} > \tau_{b}  \label{eq:Iab}
\ee

Now we do a similar thing for $I_{ba}$,
\be 
I_{ba} =\int^{\infty +i\epsilon}_{-\infty +i\epsilon}\mathcal{K}^{\gamma}(s-2\pi i \gamma) \la \mathcal{O}(\tau_{a})   \mathcal{O}(\tau_{b} +is) \ra_{\Sigma_{\gamma}},  \quad \tau_{a} >\tau_{b}. \label{eq;Ibaa}
\ee

 By shifting the contour $s \rightarrow s+\pi i \gamma$, and then flipping the sign $s \rightarrow -s$ we get

\be
I_{ba} =\f{i\sin \pi \gamma}{8\pi^2} \int^{\infty+i \epsilon}_{-\infty +i \epsilon} ds \f{-s+\pi i\gamma}{\sinh\f{s-\pi i\gamma}{2} \sinh \f{s+\pi i \gamma}{2}} G_{ab}(s-\pi i\gamma) \label{eq:Iba}
\ee

%
%
%
%
%

In the expressions (\ref{eq:Iab}) (\ref{eq:Iba}), we can take $\epsilon \rightarrow 0$.   Finally we obtain 
\be
I_{ab}+I_{ba}= \f{\gamma\sin \pi \gamma}{4\pi} \int^{\infty}_{-\infty}  \f{ds}{\sinh\f{s-\pi i\gamma}{2} \sinh \f{s+\pi i \gamma}{2}}  G_{ab}(s-\pi i\gamma) 
\ee

$T^{(2)}_{\gamma,\mathcal{O}} (\delta \rho)$ is obtained by applying the differential operator, 
\be 
T^{(2)}_{\gamma,\mathcal{O}} (\delta \rho) =C(\theta_{0}, \p_{a}) C(\theta_{0}, \p_{b})  (I_{ab}+I_{ba})
\ee

Notice that in the $\gamma \rightarrow 1$ limit,  its derivative   recovers the second order term $S^{(2)}(\delta \rho )$ entanglement entropy,
\be 
S_{\mathcal{O}}^{(2)}(\delta \rho) = C(\theta_{0}, \p_{a}) C(\theta_{0}, \p_{b})  \int^{\infty }_{-\infty}ds  \f{-1}{4\sinh^2 \left( \f{s-i \epsilon}{2}\right)} \la \mathcal{O} (is +\tau_{a}) \mathcal{O} (\tau_{b})\ra_{\Sigma_{1}}.
\ee

\section{Direct Fourier transformation} 

Here we would like to directly show that 
\begin{align}
\mathcal{K}_{n}^{\gamma}(s) &=\int^{\infty+i\epsilon}_{-\infty+i\epsilon} \f{da}{2\pi} \;  \mathcal{K}_{n}^{\gamma}(\omega) e^{-ias}  \nonumber  \\[+10 pt] 
&=\int^{\infty+i\epsilon}_{-\infty+i\epsilon} \f{da}{2\pi}\;  \f{e^{-ias}}{\sinh^2 \pi a}  \left[ (\gamma-1) -\gamma e^{2 \pi a} +e^{2\pi \gamma a} \right] 
\end{align}

The first piece is 
\be 
I_{1}=\int^{\infty+i\epsilon}_{-\infty+i\epsilon} \f{da}{2\pi} \f{e^{-ias}}{ \sinh^2 \pi a}  =\f{s}{4\pi^2} \left(\f{1}{1-e^{-s}} \right)
\ee

The second order term can be obtained by the shift $s \rightarrow s+2\pi i$, 
therefore 
\be 
I_{2}=\int^{\infty+i\epsilon}_{-\infty+i\epsilon} \f{da}{2\pi} \f{e^{-ia(s+2\pi i)}}{\sinh^2 \pi a}  =\f{(s+2\pi i) }{4\pi^2} \left(\f{1}{1-e^{-s}} \right)
\ee
Similarly, 
\be 
I_{3}= \int^{\infty+i\epsilon}_{-\infty+i\epsilon} \f{da}{2\pi} \f{e^{-ia(s+2\pi i \gamma)}}{ \sinh^2 \pi a}  =\f{(s+2\pi i \gamma) }{4\pi^2} \left(\f{1}{1-e^{-(s+2\pi i \gamma)}} \right)
\ee

Then the total integral is 
\be 
(\gamma-1) I_{1}+\gamma I_{2} + I_{3} =\f{i}{8\pi^2} \left[\f{(s+2 \pi i \gamma)  \sin \pi  \gamma}{\sinh \f{s}{2}\sinh \f{s+2\pi i \gamma}{2}}\right]
\ee

therefore we recover the first non trivial part.

\section{Details of the  holographic rewriting}
\label{sec;rewritings} 

In section \ref{subsubsec;HR}, we used the result,

\begin{align}
Y_{\gamma}(\delta \rho)&= \int dX_{B} \;\omega_{\phi} \left(K_{E}(X_{B}| \tau_{ab}, \int^{\infty}_{-\infty} ds \mathcal{Y}(s-i \epsilon) K_{R}(X_{B}|s) \right)  \nonumber \\
&= i\int dX_{B} \;  \omega_{\phi}\left( K_{E}(X_{,B}| \tau_{ba}),\; K_{E} (X_{B} |-2\pi  \gamma)-  K_{E} (X_{B} |0) \right)
\end{align}

with
\be
\mathcal{Y}(s-i \epsilon)= \f{-(\sin \pi\gamma)/4\pi }{\sinh \left(\f{s-2i\epsilon}{2}\right)\sinh \left(\f{s-2\pi i \gamma}{2}\right)}.
\ee

In this appendix, we prove this. The derivation is very similar to the one in \cite{Faulkner:2017tkh}. 

The retarded bulk to boundary propagator is given by 
\be 
K_{R}(X_{B}|s) =i \theta(s_{B} -s) \lim_{\epsilon \rightarrow 0} \left[K_{E}(X_{B}| is- \epsilon) - K_{E}(X_{B}| is+\epsilon) \right].
\ee
In particular,  as a function of $s$, the retarded propagator  is non vanishing only in the window $-\infty< s<s_{*}$.  The value of $s_{*}$ is fixed by demanding that the boundary point is null separated from the 
bulk point $X_{B}$. Then 
\begin{align}
 \int^{\infty}_{-\infty} ds \mathcal{Y}(s-i \epsilon) K_{R} (X_{B}|s) & = \int^{s_{*}}_{-\infty} \mathcal{Y}(is- \epsilon) \left[K_{E}(X_{B}| s+i \epsilon) - K_{E}(X_{B}| is+\epsilon)  \right] \nonumber \\
&=\int_{C} ds  \mathcal{Y}(s-i \epsilon)  K_{E}(X_{B}| s) ,
\end{align}
 where $C$ is the closed contour starting from $-\infty+i\epsilon$ to $s_{*}+i\epsilon$, then to $s_{*}+2(\pi-\epsilon) i$ and ending at $-\infty+2(\pi-\epsilon) i$. We also used the KMS condition $K_{E}(X_{B}|is+2\pi )=K_{E}(X_{B}|is)$,   $\mathcal{Y}(s+ 2\pi i)=\mathcal{Y}(s)$ to fix the contour.  By picking up poles of $\mathcal{Y}(s-i \epsilon)$ at  $ s= i \epsilon$ and  $ s=2\pi  i \gamma$, we obtain the result.

\bibliographystyle{utphys}
\bibliography{chaosandrelent}

\end{document}